\documentstyle[prl,aps]{revtex}
\begin{document}
\title{Interplay between charge, orbital and magnetic order in 
Pr$_{1-x}$Ca$_x$MnO$_3$}
\author{M. v. Zimmermann$^1$, J.P. Hill$^1$, Doon Gibbs$^1$, M. Blume$^1$, 
D. Casa$^2$,
B. Keimer$^{2,3}$, Y. Murakami$^4$, Y. Tomioka$^5$, and Y. Tokura$^6$}

\address{$^1$Department of Physics, Brookhaven National Laboratory, Upton, New 
York 11973, USA}

\address{$^2$Department of Physics, Princeton University, New Jersey 08544, USA}

\address{$^3$Max-Planck-Institut f\"ur Festk\"orperforschung, 70569, Stuttgart, 
Germany.}

\address{$^4$Photon Factory, Institute of Materials Structure Science, High 
Energy Accelerator Research Organization, Tsukuba, 305-0801, Japan}

\address{$^5$Joint Research Center for Atom Technology (JRCAT), 
Tsukuba 305-0046, Japan}

\address{$^6$Department of Applied Physics, University of Tokyo, Tokyo 113-0033, 
Japan and JRCAT}

%
\maketitle
\begin{abstract}
We report resonant x-ray scattering studies of charge and orbital order
in Pr$_{1-x}$Ca$_{x}$MnO$_3$ with $x$=0.4 and 0.5. Below the
ordering temperature, T$_O$=245 K,
the charge and orbital order intensities follow the same temperature
dependence, including an increase at the antiferromagnetic
ordering temperature, $T_N$.  
High resolution measurements reveal, however, 
that long range orbital order is never achieved. 
Rather, an orbital domain state is formed.
Above T$_O$, the
charge order fluctuations are more highly correlated than
the orbital fluctuations. Similar phenomenology is observed
in a magnetic field.
We conclude that the charge order drives the orbital order
at the transition.
\end{abstract}
\narrowtext
\twocolumn
%

Disentangling the origins of high temperature superconductivity and colossal 
magnetoresistance in the transition metal oxides remains at the center of 
current activity in condensed matter physics. An important aspect of these 
strongly correlated systems is that no single degree of freedom dominates their 
response. Rather, the ground state properties are thought to reflect a balance 
among several correlated processes, including orbital and charge order, 
magnetism, and coupling to the lattice.

The perovskite manganites provide an especially illuminating example of 
the interplay among these
interactions, since in these materials the balance may be altered by doping, 
among other things. As a
result, much work has been done to understand their magnetic ground states 
and lattice distortions,
dating back to the seminal experiments of Wollan and Koehler \cite{Wol55}.
Less is known about the roles of charge and orbital order, 
for which (until recently) there has been no direct experimental probe. The 
classic work of Goodenough \cite{Goo55} has nevertheless served as a guide to 
their ordered 
arrangements, as supplemented by measurements of the temperature dependence of
the lattice constants, and other properties. Recently this situation has changed
with the detection of charge and orbital order by x-ray resonant scattering 
techniques. 
\cite{Mur98a,Mur98b,Ishihara-PRL-1998,Fabrizio-PRL-1998,Sawatzky-preprint,Endoh-preprint,Paolasini-preprint,Benfatto-preprint}. 
Specifically, it has been found that
the sensitivity of x-ray scattering to these structures is dramatically
enhanced by tuning the incident x-ray energy to the Mn K-absorption edge. Thus,
it is now possible to characterize the orbital and charge order distributions 
directly, and to study their response to changes of temperature or magnetic 
field.

In this paper, we report x-ray resonant scattering studies of 
Pr$_{1-x}$Ca$_x$MnO$_3$ with 
$x$=0.4 and 0.5. We have detected both charge and orbital order 
below a common phase transition temperature (T$_O$=245 K), and confirmed the 
ground state originally proposed by Goodenough for the isostructural compound 
La$_{0.5}$Ca$_{0.5}$MnO$_3$ \cite{Goo55}. Below the transition, we find that the
intensities of the charge and orbital order have the same temperature 
dependences,
suggesting that they are linearly coupled. There is, moreover, a notable
increase of the scattering at the N\'eel temperature (T$_N$=170 K), implying a
coupling of the orbital and magnetic degrees of freedom. Intriguingly, high 
momentum
transfer resolution measurements reveal that long range orbital order is never
achieved at these concentrations. 
Rather, a domain state is formed at low temperature. 
At temperatures
above T$_O$, we observe critical charge and orbital scattering. 
Remarkably, the correlation lengths differ, with the length scale of the charge 
order exceeding that of the 
orbital order. From this we conclude that charge order drives the orbital
order in these systems. This picture is supported by studies in which the phase 
transition is driven by an applied magnetic field.

For small $x$, Pr$_{1-x}$Ca$_x$MnO$_3$ has an orbitally ordered ground state at 
low temperature analogous to that observed in LaMnO$_3$. 
The electronic configuration of the Mn$^{3+}$ ions 
is ($t_{2g}^3$,$e_g^1$), with the $t_{2g}$ electrons localized. The $e_g$ 
orbitals are hybridized with the oxygen $p$ orbitals, and participate in a 
cooperative Jahn-Teller distortion of the MnO$_6$ octahedra. This leads to 
(3$x^2-r^2$)-(3$y^2-r^2$)-type orbital order of the $e_g$ electrons in the 
$ab$-plane. For $0.3\le x \le 0.7$, charge 
order among Mn$^{3+}$ and Mn$^{4+}$ ions is believed to 
occur in addition to the orbital order. The fraction of Mn ions in the Mn$^{4+}$
state equals the concentration of Ca ions. Thus, by varying the Ca 
concentration, it is
possible to alter the balance between the charge and orbital order. 
The proposed ground
state for $x=$0.4, including charge, orbital and magnetic order, is illustrated 
in fig. 1b \cite{Goo55,Jir85,Okimoto}.
In orthorhombic notation, for which the fundamental Bragg peaks occur at (0,2k,0)
with k integer, the charge order reflections occur at (0,2k+1,0) and the orbital
order reflections at (0, k+$\frac{1}{2}$,0).  The magnetic structure
is of the modified CE-type \cite{Jir85}.

The single crystals used in this study were grown by a float zone
technique. (0,1,0) surfaces were cut and polished giving mosaic spreads of 
$\sim 0.25^\circ$.
X-ray scattering experiments were carried out at the National Synchrotron Light 
Source
on Beamlines X22B and C. X22C utilizes a vertical scattering geometry with 
an incident energy resolution of $\sim$ 5 eV. Two analyzer configurations were 
used. The first, a Ge(111) crystal, provided an effective resolution
of 7.2 $\times 10^{-4} \AA^{-1}$  (HWHM) at the (0,2,0) reflection. The second 
provided linear polarization analysis of the
scattered beam with a Cu (220) crystal \cite{Gibbs-poln}. The incident 
polarization was 95\% linearly polarized in the horizontal plane ($\sigma$). 
Experiments in a magnetic field were performed on X22B in a horizontal 
scattering geometry.

The present experiments were carried out using x-ray resonance scattering 
techniques. As shown in a series of recent papers 
\cite{Mur98a,Mur98b,Ishihara-PRL-1998,Fabrizio-PRL-1998,Sawatzky-preprint,Endoh-preprint,Paolasini-preprint,Benfatto-preprint},
scattering from orbital and charge 
order in transition metal oxides is enhanced when the incident x-ray energy is 
tuned near the K absorption edge. In the dipole approximation, this corresponds 
to a $1s \rightarrow 4p$ transition of a virtually excited electron at
the metal site. The sensitivity to orbital order in manganites arises from the
splitting of the Mn 4p levels by the 3d levels, which is mainly associated
with the Jahn-Teller distortion {\cite{Sawatzky-preprint,Benfatto-preprint}. The 
sensitivity to
charge order originates in the small difference in K-absorption energies 
associated with the Mn$^{3+}$ and Mn$^{4+}$ sites, leading to anomalous 
scattering at the difference reflections \cite{Mur98a}.

The inset of fig. 1 shows the energy dependence of the intensities of the 
charge and orbital order as the incident photon energy is tuned through the K
absorption edge. The data were taken at the (0,3,0) and (0,2.5,0) reflections,
respectively, of the $x=$0.4 sample. Each scan shows an enhancement at 6.555 keV,
characteristic of dipole resonant scattering. Maximum count rates of 
about 800 and 3000 s$^{-1}$
were obtained for the orbital and charge order scattering, respectively, with 
a Ge(111) analyzer. The structure observed in the lineshape for the charge order
reflects the interference of the resonant and nonresonant contributions, and 
will be discussed in detail elsewhere \cite{Zimmermann-tobe}. In addition to 
the resonance, both the charge and orbital
order intensities exhibited a $\sin^2(\psi)$ dependence on the azimuthal angle,
$\psi$, at resonance \cite{Mur98a,Mur98b}.
This angle
defines rotations of the sample about the scattering vector.
Polarization analysis further revealed that the orbital scattering is 
predominantly
rotated ($\sigma$-$\pi$), whereas the charge scattering is mainly unrotated
($\sigma$-$\sigma$). These results are consistent with predictions of the 
resonant cross-section for orbital and charge order 
\cite{Mur98a,Mur98b,Ishihara-PRL-1998},
and confirm the original picture of charge and orbital order
given for this class of compounds by Goodenough \cite{Goo55} (fig. 1b).

The temperature dependences of the charge and orbital order obtained at resonance
for the $x=$0.4 sample are shown in fig. 1. For comparison purposes  the 
intensities have been scaled together at 10 K. Between 10 and 120 K, both 
intensities are approximately constant, but decrease by about 25\% on passing 
through the N\'eel 
temperature (T$_N$=170 K). They drop sharply to zero at T$_O$=245 K. This is 
coincident with an orthorhombic-to-orthorhombic structural transition, as 
determined from high-resolution measurements of the lattice constants 
\cite{Zimmermann-tobe}. It follows that the temperature dependences of the 
charge and orbital order are identical, which suggests that 
the corresponding order parameters are linearly coupled. It seems clear
in this regard that the growth of orbital (and charge) order below 245 K 
enhances the antiferromagnetic correlations, and thereby promotes the magnetic 
phase transition. This is consistent with the results of inelastic 
neutron scattering studies of (Bi,Ca)MnO$_3$,
in which orbital order was found to quench ferromagnetic 
fluctuations \cite{Bao97}.
Qualitatively similar results have been found for the $x=$0.5 sample
\cite{Zimmermann-tobe}.

The behavior of the charge and orbital scattering in the vicinity of the 
structural phase transition at T$_O$ is illustrated for the $x=$0.4 sample in 
fig. 2. Longitudinal scans were taken (upon warming) of the (0,3,0) reflection 
in a $\sigma$-$\sigma$ geometry and of the (0,2.5,0) reflection in 
a $\sigma$-$\pi$ geometry. We find that a measurable intensity of the charge 
order fluctuations (shown on a log scale in fig. 2a) persists 
to much higher temperatures above T$_O$ 
than the orbital order fluctuations.
The corresponding peak widths are considerably narrower for the charge order
(fig. 2b), implying
that the correlation lengths of the charge order are longer than
those of the orbital order at any given temperature above T$_O$ \cite{size-note}.

The picture these data then present is one in which the phase transition
proceeds via local charge order fluctuations which grow as the
transition is approached, nucleating long range order at the transition 
temperature.
The orbital fluctuations are induced by these charge fluctuations through
the coupling discussed above, and become observable only close to the
transition. 

The  phase transition may also be
driven by applying a magnetic field,
as demonstrated by Tomioka {\em et al.}
\cite{Tom96}. It is an interesting question
whether the same phenomenology of the fluctuations applies
when the temperature is held fixed. We have carried out studies of the
transition at two temperatures, T=30 K and 200K, with critical
fields of H$_O$=6.9(1) T, and H$_O$=10.4 T, respectively.
Data taken at T=30 K are illustrated in fig. 3.
The two order
parameters exhibit identical field dependences below the transition.
Above the transition, the charge order fluctuations are markedly stronger
than the orbital fluctuations. Similar behavior was observed at
T=200 K, i.e. charge order fluctuations were observed
at fields  for which orbital fluctuations were no longer
observable (inset fig. 3). Thus, it appears that the
transition is driven by charge order fluctutations for both temperature
and field driven cases. (As a result of experimental constraints, it
was only possible to measure charge and orbital order
scattering at a photon energy of 8 keV in an applied magnetic field. The 
corresponding nonresonant intensities
are sufficiently weak above T$_O$ that it was not possible to obtain
reliable values of the half-width).

Finally, we performed high q-resolution measurements of the
charge and orbital order below T$_O$
to investigate the extent of their order in detail.
Remarkably, in the $x$=0.4 sample, we found an orbital order correlation length 
of $\xi_{OO} = 320\pm10$ \AA{} and a charge order correlation length of 
$\xi_{CO} \geq 2000 $ \AA.
This difference is even more apparent in the $x$=0.5 sample, as
shown in fig. 4. Here longitudinal scans through the  (0,2,0), (0,1,0)
and (0,2.5,0) reflections are superimposed for comparison.
The width of the (0,2,0) scan approximates the effective resolution. 
The (0,1,0) charge order reflection shows only a  slight broadening
(9.1(1) $\cdot$ 10$^{-4}$ \AA$^{-1}$), corresponding to a correlation length of
$\xi_{CO} \geq 2000$ \AA{} \cite{note-res}.
The orbital order, however, is substantially broadened, with a correlation 
length of $\xi_{OO} = 160 \pm 10$ \AA \cite{footn1}.
It follows that the orbitals do not exhibit long range order, but instead
form a domain state with randomly distributed anti-phase domain
walls (see fig. 1b). In contrast, the charge order is much more highly 
correlated.

A possible explanation for the difference in orbital domain sizes
observed in the two samples follows from the fact 
that the $x=$0.5 sample is closer
to tetragonal than the $x$=0.4 sample:
$\delta(x=0.5)=\frac{2(a-b)}{(a+b)}=1.48\times10^{-3}$
compared to $\delta(x=0.4)= 4.23 \times 10^{-3}$ at room temperature
\cite{Jir85}. In the more tetragonal sample the $a$ and
$b$ domains are nearly degenerate and the energetic cost of an orbital
domain wall is correspondingly reduced \cite{Millis-private}.

The presence of an orbital domain state is consistent with powder
neutron diffraction studies of
$\rm La_{0.5}Ca_{0.5}MnO_3$, which also exhibits the CE magnetic 
structure with orbital and charge order
\cite{Rad97}. In this material, magnetic correlation lengths of
$\xi_{3+}=200-400$ \AA{} and 
$\xi_{4+}\geq 2000$ \AA{} 
were found for the respective sublattices, and anti-phase
domain walls were postulated
to explain the disorder.  It seems likely that these domain walls are
in fact
the orbital domain walls observed in the present experiment, which 
lead to magnetic disorder through the coupling mentioned above.
The presence of such orbital domain boundaries breaks the
magnetic coherence of the $3+$ sublattice only, as long as charge order
is preserved (fig. 1b). These  
results taken together suggest that orbital domain states may be common 
to these systems -- at least in this range of doping.

In summary, we have used resonant x-ray scattering techniques to study
the development of charge and orbital order in doped
manganites, Pr$_{1-x}$Ca$_x$MnO$_3$, with 
$x$=0.4 and 0.5. We have found that the transition into
a charge and orbitally ordered state proceeds via charge order
fluctuations, which grow as the transition is approached from above,
until at an abrupt, first-order-like, transition long range charge
order is nucleated. At the same time, less well correlated orbital
fluctuations are observed. While the correlation length of the orbital order
grows as the transition is approached, long range order is not
achieved, and below the charge order transition an orbital domain
state is observed. The orbital correlation length appears to be
concentration dependent, with a more highly correlated orbital state
being observed in the $x$=0.4 sample. Phenomenologically similar behavior
is observed when the phase transition is driven by
a magnetic field. Below the transition, the charge and
orbital order parameters exhibit identical temperature dependences,
indicative of a linear coupling between these degrees of freedom. At the
N\'eel temperature a jump in their intensities is observed
demonstrating the importance of the magnetic -charge/orbital order
coupling in these systems.

We acknowledge useful conversations with A.J. Millis and
G.A. Sawatzky.
The work at Brookhaven was supported by the U.S. Department of Energy,
Division of Materials Science, under Contract No. DE-AC02-98CH10886 and 
at Princeton University by the N.S.F., under grant DMR-9701991.
Support from the Ministry of Education, Science and Culture, Japan, by
the New Energy and Industrial Technology Development Organization (NEDO)
and by the Core Research for Evolutional Science and Technology (CREST)
is also acknowledged.

%

\begin{figure}
\label{fig1}
\caption{a: Orbital and charge order parameters versus temperature for the $x=$.4
sample.  Open and closed circles: Resonant intensities of the orbital and charge 
order measured at the (0,1.5,0) and (0,3,0) reflections, respectively. Inset:
Energy dependence of the orbital and charge order peaks in the $\sigma$-$\pi$
and $\sigma$-$\sigma$ geometries, respectively. b: schematic of charge and 
orbital order, showing an orbital anti phase domain boundary (dashed line) in 
the $ab$-plane. Black points represent Mn$^{4+}$ and open symbols Mn$^{3+}$. The 
arrows indicate the magnetic ordering. The dotted and solid lines show the 
charge and orbital unit cells, respectively.}
\end{figure}

\begin{figure}
\label{fig2}
\caption{a) Temperature dependence of the peak intensities of the charge 
(closed circles) and orbital (open circles) for the $x=$0.4 sample.
b) Temperature dependence of the half widths at half maximum (HWHM).}
\end{figure}

\begin{figure}
\label{fig4}
\caption{Charge and orbital order as function
of magnetic field at 30 K. The intensities of the two peaks have been scaled 
to agree at low fields. 
Inset: Charge and orbital order
superlattice reflection at 198 K and 11 T. The orbital order is no
longer observable, but scattering from charge order remains clearly
visible.}
\end{figure}

\begin{figure}
\label{fig3}
\caption{a: Longitudinal scans of the (0 2 0) Bragg reflection, the (0 1 0)
charge order peak and the (0 2.5 0) orbital order peak.}
\end{figure}


\begin{thebibliography}{}
\bibitem{size-note}
The correlation length of the charge order must be at least as long  
as that of the orbital order, since the unit cell of the orbital order is defined
on the charge order lattice.
\bibitem{note-res}
As estimated by performing a 1-d deconvolution of the resolution
and using $\xi=\frac{b}{2\pi \Delta k}$, were $\Delta k$ is
the fitted HWHM.
\bibitem{footn1}
Note that although the resolution function exhibits some q-dependence, 
the observed broadening is much too large to
be explained by such effects.
\end{thebibliography}
\end{document}